\documentclass[%
 reprint, amsmath,amssymb, aps,showpacs]{revtex4-1}

\usepackage{graphicx}
\usepackage{epsfig}
\begin{document}


\title{Chiral magnetic effect and Maxwell-Chern-Simons electrodynamics  \\
in Weyl semimetals}


\author{Debanand Sa}%
\affiliation{%
Department of Physics, Banaras Hindu University, Varanasi-221 005
  \\}%
  
\date{\today}

\begin{abstract}
The Weyl semimetal, due to a non-zero energy difference in the pair of Weyl nodes shows chiral magnetic effect(CME). This leads to a flow of dissipationless electric current along an applied magnetic field. Such a chiral magnetic effect in Weyl semimetals has been studied using the laws of classical electrodynamics. It has been shown that the CME in such a  semimetal changes the properties namely, frequency dependent skin depth, capacitive transport, plasma frequency etc. in an unconventional way as compared to the conventional 
metals. In the low frequency regime, the properties are controlled by a natural length 
scale due to CME called the chiral magnetic length. Further, unlike the conventional metals, the plasma frequency in this case is shown to be strongly magnetic field-dependent. Since the plasma frequency lies below the optical frequency, the Weyl semimetals will look transparent. Such new and novel observations might help in exploiting these class of materials in potential applications which would completely change the future technology.
\end{abstract}

\maketitle



In recent times, there has been an explosion of research in topologically 
non-trivial states of matter following the remarkable discovery of topological 
insulators and superconductors\cite{kane05,mele05,bernevig06,moore07,konig07,xia09,hasan10,qi11}. The topological order menifested 
in these systems is not associated with the spontaneously breaking of a symmetry rather 
can be described by topological invariants. Usually the robust topological protection is associated with a non-zero spectral gap in the bulk and the existence of protected zero 
energy surface states is regarded as the hallmark of a non-trivial topological phase of matter. However, recently there has been proposal that systems in three spatial dimensions 
in presence of broken time reversal(TR) symmetry/space inversion(SI) symmetry can also be topologically protected even without bulk energy gap\cite{wan11,yang11,burkov11,xu11,zyuzin12a,zyuzin12b,meng12,gong11,sau12}. 
These are called Weyl semimetals(WSM)\cite{wan11}. 
Realization of such phenomena has been predicted in pyrochlore iridates\cite{wan11} 
and since then, it has been reported to be experimentally confirmed in materials such 
as,TaP, NbP, TaAs and NbAs\cite{xu15a,lv15a,lv15b,shekhar15,yang15,xu15b}. 

A WSM is a three-dimensional analogue of graphene and the low energy excitations with 
broken TR symmetry can be described by a pair of linearly dispersing massless Dirac 
fermions governed by the Hamiltonian, $H_{\chi}(k)=\chi\hbar v_F \vec{k}.\vec{\sigma}-\mu_{\chi}$, where $\chi=\pm$ is the chirality, $v_F$, the Fermi velocity, 
$\sigma=(\sigma_x, \sigma_y, \sigma_z)$ refers to three Pauli matrices and $\mu_{\chi}$ 
stands for chirality dependent chemical potential(given by a superposition of the 
equilibrium carrier density and the pumped carrier density originating from chiral anomaly). 
Thus, here $H_{\chi}$ describes the two Weyl fermions of opposite chirality. Due to the fermion doubling theorem\cite{nielsen83}, Weyl nodes with opposite chirality always appear 
in pairs. The two band touching Weyl points act as a source and a sink(monopole and antimonopole charges) of Berry curvature which is similar to a fictitious magnetic field on the electron wave function in momentum space\cite{xiao06}. The Weyl nodes can be separated  by a wave vector $Q$ in the first Brillouin zone or by an energy offset $\hbar Q_0$. The toplogical properties of a Weyl semimetal are menifested in the form of a $\theta$-term contribution to the action $S_{\theta}=\frac{\alpha}{4\pi^2}\int dt\int d^3 r \theta(\vec{r},t)\vec{E}.\vec{B}$ where $\theta(\vec{r},t)=2(\vec{Q}.\vec{r}-Q_0 t)$ is so called axion angle and 
$\alpha=\frac{e^2}{\hbar c}\approx 1/137$ is the fine structure constant. Here, $e$ is the electron charge and $\vec{E}$,$\vec{B}$ are respectively the electric and magnetic fields. 
If the bands are degenerate with $\vec{Q}=0=Q_0$, the system does not posses topological properies, that is, such effect would disappear if the two nodes with opposite monopole charges merge and annihilate each other. Due to the non-trivial topology in the momentum space(see  Fig. 1(a)), such materials exhibit a wide variety of unusual electromagnetic responses\cite{hosur13,burkov15}. In a Weyl semimetal, 
electrons near each Weyl node can be associated with a chirality by the monopole charge of that node. In the application of a pair of non-orthogonal electric and magnetic fields, the charges can be transported between the two Weyl nodes with opposite chiralities. Thus, the number of electrons with a definite chirality is no longer conserved implying the so called chiral anomaly\cite{adler69} in these systems. Such anomaly is predicted to give an enhanced negative magnetoresistence\cite{son13} when the applied electric and magnetic fields are parallel to each other. Such predictions have been confirmed by experiments in the material TaAs\cite{huang15,zhang15}. 

\begin{figure}
\includegraphics[scale=0.40]{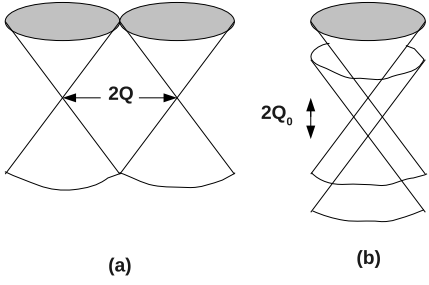} 
\caption{\label{fig33} Schematic diagram of two nodes in a Weyl semimetal separated in (a) momentum ($2\vec{Q}$) and (b) energy ($2Q_0$) space.}
\end{figure} 

In addition, these 
nodal materials might show chiral magnetic effect(CME) when the energies of the pair of 
Weyl nodes are different\cite{zyuzin12a,zyuzin12b,goswami13,step12,fuku08}(see Fig. 1(b)). This effect provides a dissipationless electric current $\vec{J}_{ch}$ 
flowing along an applied magnetic field $\vec{B}$($\vec{J}_{ch}=\sigma_{ch}\vec{B}$) in 
contrast to the conventional Ohm's law($\vec{J}=\sigma_D\vec{E}$). The CME conductivity 
$\sigma_{ch}$, in a low energy effective theory is shown to be proportional to the energy 
separation between the pair of Weyl nodes\cite{burkov15}. However, at present there is no direct observation of CME in its original sense, that is, an electric current driven by the external magnetic field. Moreover, CME vanishes as a static effect in the thermal equilibrium\cite{vazifeh13,chen13} whereas it might exist in the non-equilibrium limit\cite{goswami13,chang15}. Since CME can be regarded as a Chern-Simons extension of 
electromagnetism\cite{griffiths} which includes electric and magnetic fields, electric 
charge and current densities, such phenomena need investigation within the framework of 
the laws of electrodynamics. This might lead to non-trivial consequences in many of their properties. 

In this work, we start with Maxwell-Chern-Simons(MCS) equations of electrodynamics and show 
the effect of CME in the properties such as, frequency dependent skin depth, capacitive transport, plasma frequency etc. in a Weyl semimetal. Further, these properties which look unconventional are compared to that of conventional metals. Such  non-trivial observations 
might be exploited for their applications in future electronics.  

The axion term modifies the standard Maxwell equations of electrodynamics in a Weyl semimetal. Thus, the MCS equations in a WSM can be written as, 

\begin{eqnarray} 
\vec{\nabla}\cdot\vec{E}=4\pi (\rho_{+} +\rho_{-}) \\
\vec{\nabla}\cdot\vec{B}=0 \\
\vec{\nabla}\times\vec{E}=-\frac{1}{c}\frac{\partial\vec{B}}{\partial t} \\
\vec{\nabla}\times\vec{B}=\frac{4\pi}{c}(\vec{J}_{+}+\vec{J}_{-})
+\frac{1}{c}\frac{\partial\vec{E}}{\partial t},
\end{eqnarray} 

\noindent where the source terms respectively are both the charge densities $\rho_{\pm}$ and   both the current densities $\vec{J}_{\pm}$ in the two valleys(two chiralities). The modifications in the source terms(Eq.(1) and (4))\cite{dsa1} are due to the axion action discussed in the introduction. The fluctuations in the charge density in both the valleys are related to the current fluctuations via the continuity equation. For the Weyl semimetal, it contains an anomalous non-vanishing contribution due to chiral anomaly. The continuity equation reads as, 

\begin{equation} 
\frac{\partial\rho_{\pm}}{\partial t} +
\vec{\nabla}\cdot\vec{J}_{\pm}=\pm\frac{e^3}{4\pi^2}\vec{E}\cdot\vec{B}.
\end{equation} 

\noindent The equation for the current includes at least two terms in the limit 
$q\rightarrow 0$(neglecting the diffusion current); the conventional Drude contribution and the chiral current which flows along the magnetic field. Thus, the equation for the current is written as, 

\begin{equation} 
\vec{J}_{\pm}=\frac{\sigma_D}{2}\vec{E}\pm\frac{e^2}{4\pi^2}\mu_{0\pm}\vec{B}, 
\end{equation} 

\noindent where $\sigma_D$ is the Drude conductivity and $\mu_{0\pm}$ are the chemical potentials in the two valleys. The fluctuations in the chemical potential are related to the fluctuations in the charge density in each valley via the usual Thomas-Fermi relation as, 

\begin{equation} 
\rho_{\pm}=e g_B(\mu_{0\pm}-\epsilon_F), 
\end{equation} 

\noindent where $\epsilon_F$ is the unperturbed Fermi energy of the system and $g_B$ is 
the density of states at $\epsilon_F$ which can depend on the magnetic field. 

Due to the coupling of $\vec{E}$ and $\vec{B}$ in the MCS equations, one can go for higher partial derivatives to decouple them. Thus, the Faraday's and the Ampere's law(Eq.(3) and (4))  yields, 

\begin{eqnarray} 
{\nabla}^2\vec{E}=\frac{4\pi}{c^2}[\sigma_D\frac{\partial\vec{E}}{\partial t}
+ \sigma_{ch}\frac{\partial\vec{B}}{\partial t}] 
- \frac{1}{c^2}\frac{\partial^2\vec{E}}{\partial t^2}\\
{\nabla}^2\vec{B}=\frac{4\pi}{c^2}\sigma_D\frac{\partial\vec{B}}{\partial t} 
+\frac{1}{c^2}\frac{\partial^2\vec{B}}{\partial t^2} 
-\frac{16\pi^2}{c^2}\sigma_{ch}^2\vec{B} \nonumber\\-\frac{16\pi^2}{c^2}\sigma_D\sigma_{ch}\vec{E}
-\frac{4\pi}{c^2}\sigma_{ch}\frac{\partial\vec{E}}{\partial t},
\end{eqnarray} 

\noindent where $\sigma_{ch}=\frac{e^2}{4\pi^2}(\mu_{0+}-\mu_{0-})=\frac{e^2}{4\pi^2}\frac{(\rho_{+}-\rho_{-})}{e g_B}$. Using the Fourier mode expansion of both the fields,  $\vec{E}(\vec{r},t)=\vec{E_0}\exp{[i(\vec{k}.\vec{r}-\omega t)]}$ and $\vec{B}(\vec{r},t)=\vec{B_0}\exp{[i(\vec{k}.\vec{r}-\omega t)]}$, Eqs.(8) and (9) can be 
simplified as, 

\begin{eqnarray} 
(k^2-\frac{\omega^2}{c^2}-i\frac{4\pi}{c^2}\sigma_D\omega)\vec{E_0}=(i\frac{4\pi}{c^2} \sigma_{ch}\omega)\vec{B_0}\\
(k^2-\frac{\omega^2}{c^2}-i\frac{4\pi}{c^2}\sigma_D\omega-\frac{16\pi^2}{c^2}\sigma_{ch}^2)\vec{B_0}\nonumber \\ 
=(\frac{16\pi^2}{c^2}\sigma_D\sigma_{ch}-i\frac{4\pi}{c^2}\sigma_{ch}\omega)\vec{E_0}.
\end{eqnarray} 

\noindent Thus, the ratio between the amplitudes of both the fields can be written as, 

\begin{eqnarray} 
\frac{{B_0}}{{E_0}}=\frac{(k^2-\frac{\omega^2}{c^2}-i\frac{4\pi}{c^2}\sigma_D\omega)}{(i\frac{4\pi}{c^2} \sigma_{ch}\omega)} \nonumber \\
=\frac{(\frac{16\pi^2}{c^2}\sigma_D\sigma_{ch}-i\frac{4\pi}{c^2}\sigma_{ch}\omega)}{(k^2-\frac{\omega^2}{c^2}-i\frac{4\pi}{c^2}\sigma_D\omega-\frac{16\pi^2}{c^2}\sigma_{ch}^2)}.
\end{eqnarray} 

\noindent Such an equation yields the dispersion relation(the relation between $k$ and $\omega$) in the form of an algebraic equation as, 

\begin{equation} 
{(k^2-\frac{\omega^2}{c^2}-i\frac{4\pi}{c^2}\sigma_D\omega)}^2
={(\frac{4\pi}{c}\sigma_{ch} k)}^2.
\end{equation}   

\noindent This equation can be rewritten in the SI units as, 

\begin{equation} 
{(k^2-\mu\epsilon\omega^2-i\mu\sigma_D\omega)}^2
={(k\mu\sigma_{ch})}^2,
\end{equation}   

\noindent where $\mu$ and $\epsilon$ are respectively the magnetic permeability and the dielectric permittivity of the material. The above equation contains four roots, given as, 

\begin{equation} 
k=\frac{1}{2}[{\pm\mu\sigma_{ch}}\pm\sqrt{\mu^2\sigma_{ch}^2 +4 (\mu\epsilon\omega^2+i\mu\sigma_D\omega)}].
\end{equation}   

\noindent Defining a small dimensionless parameter as, $\delta=\frac{\sigma_D\omega}{\mu\sigma_{ch}^2}<<1$ and assuming a good conductor limit, that is, $\frac{\epsilon\omega}{\sigma_D}<<1$, the above dispersion(for positive sign only) can be approximated as, 

\begin{eqnarray} 
k\simeq \begin{cases}\mu\sigma_{ch}[1+\delta\frac{\epsilon\omega}{\sigma_D}+i\delta] \\
-i\mu\sigma_{ch}\delta[1+ i\frac{\epsilon\omega}{\sigma_D}] \end{cases}
\end{eqnarray}




\noindent From these equations, it is obvious that $k$ is in general complex which has 
real and imaginary parts. Writing $k=k_R\pm i k_I$, one obtains, 
$k_R=\mu\sigma_{ch}[1+\delta\frac{\epsilon\omega}{\sigma_D}]\simeq\mu\sigma_{ch}$ and $k_I=\mu\sigma_{ch}\delta$. Since the imaginary part of $k$ results in the attenuation 
of the electromagnetic wave, the so called {\it skin depth} can be calculated to be, $d=\frac{1}{k_I}={(\mu\sigma_{ch}\delta)}^{-1}$. Defining a characteristic length scale in the system as the {\it chiral magnetic length} $\l_{ch}=\frac{\pi}{\mu\sigma_{ch}}$, 
the skin depth can be rewritten as, $d=\frac{\l_{ch}}{\pi\delta}$. Thus $\l_{ch}$ gives a measure of how far the wave penetrates into the Weyl semimetal. It is obvious from this expression that the skin depth in a Weyl semimetal is dependent on frequency through the parameter $\delta$ which is quite different from a conventional metal\cite{dsa2}. 
Considering the parameters from a lattice model\cite{tewari13}, that is, taking the 
energy difference between the Weyl nodes to be $\Delta\sim 100 [meV]$, $\sigma_{ch}\sim 10^7\Delta[A/m^2 T]=10^{9}[A/m^2 T]$,  the perturbative parameter $\delta$ comes out to 
be $\delta=10^{-1}$ for a frequency $\omega=10^5[Hz]$ and 
$\frac{\epsilon\omega}{\sigma_D}=10^{-12}<<1$. The validity of these approximations become better and better for frequency less than $10^5[Hz]$. Since $\l_{ch}\sim 10^{-3}$, the 
skin  depth in such a case is estimated as, $d\sim 10^{-2}[m]$ whereas, for a normal metal 
$d\sim 10^{-3}[m]$. It is obvious that the skin depth in a Weyl semimetal is one order of magnitude larger as compared to that of conventional metal. Here, the parameters considered are, $\sigma_D\sim 10^6[S/m]$, $\mu=10^{-6}[N/A^2]$, $\epsilon=10^{-11}[C^2/N.m^2]$. The real part of $k$, on the contrary, determines the propagation speed, wavelength and the index of refraction respectively 
as, $v=\frac{\omega}{k_R}$, $\lambda=\frac{2\pi}{k_R}$ and $n=\frac{c k_R}{\omega}$.
 
\noindent Putting the expression of $k$ from Eq.(16) in Eq.(12), one obtains, 

\begin{equation} 
\frac{B_0}{E_0}\approx\frac{i\mu\sigma_{ch}}{\omega}\approx \frac{i\pi}{\omega\l_{ch}}.
\end{equation}

\noindent It is obvious from this expression that the natural length  $\l_{ch}$ 
characterizes the relation between both the fields $E_0$ and $B_0$. Writing down the wave 
number as well as both the fields in terms of their magnitudes and phases as, $k=k\exp{[i\phi]}$, $B_0=B_0\exp{[i\phi_B]}$ and $E_0=E_0\exp{[i\phi_E]}$, one can obtain 
the relation between the phases of both the fields as, $\phi_B-\phi_E\approx\phi\approx\pi/2$. This implies that the magnetic and the electric fields are no longer in phase, magnetic field 
lags behind the electric field by a phase of $\pi/2$. This can be understood in a way that the current in such a system leads the voltage by the phase of $\pi/2$, implying that the response is purely capacitive. This result is due to purely CME effect. This is in 
accordance with a recent prediction on capacitive transport\cite{fujita16} in a WSM. 
Also, under certain conditions, this might lead to material independent universal 
effective capacitance in a cylinder geometry\cite{fujita16}. Further, Eq.(17) also 
justifies the perturbative parameter $\delta$ (for $\omega\leq 10^5[Hz]$) which is the 
ratio between both the fields with their respective conductivities, that is, $\delta\approx\mid\frac{\sigma_D E_0}{\sigma_{ch}B_0}\mid=\mid\frac{i\sigma_D\omega}{\mu\sigma_{ch}^2}\mid$. This is nothing but the ratio between the ordinary current to 
that of the CME one and the analysis physically corresponds to a large CME expansion. 
Since the operational frequency here is in the radio frequency range, these materials might be applicable in radio electronics.

Next, let us consider the high frequency behaviour of WSM which can provide us a crude estimate of the plasma frequency in such systems. This can be achieved by solving the set of Eqs.(5-7) simultaneously with the help of Maxwell equations. Thus, the equation for the difference in charge density fluctuations yields, 

\begin{equation} 
\omega^2+4\pi i\sigma_D\omega-\frac{8\pi B^2}{g_B}(\frac{e^2}{4\pi^2})^2=0.
\end{equation} 

\noindent The solution of this equation provides plasma frequency which is imaginary in the case $B=0$. Such a relation can be made meaningful by replacing the Drude conductivity by its AC analogue, that is, $\sigma_D(\omega)=\frac{\sigma_D}{1-i\tau\omega}$, ($\sigma_D=\frac{ne^2\tau}{m}$ for conventional metals but in undoped WSM, $\sigma_D=\frac{\mu_0^2 e^2\tau}{12\pi^2 v_F}$ and in doped WSM $\sigma_D=\frac{e^2\tau}{24\pi^2 v_F}(\mu_{0+}^2+\mu_{0-}^2$), $\mu_0$ being the chemical potential of the nodal metal whereas $\mu_{0\pm} $ are the chiral chemical potential in WSM)\cite{tabert13}). Considering the collisionless limit $\omega\tau>>1$, the conductivity in a conventional metal becomes, $\sigma_D(\omega)=-\frac{ne^2}{im\omega}$ whereas taking the same limit in a WSM, one obtains, 

\begin{equation} 
\omega^2=\omega_p^2=\frac{4\pi\sigma_D}{\tau} +\frac{8\pi B^2}{g_B}(\frac{e^2}{4\pi^2})^2, 
\end{equation} 
 
\noindent where $\omega_p$ is the plasma frequency. In principle, the magnitude of the 
Drude conductivity and the density of states depend on the magnetic field $B$ and hence can 
be determined from the fact that how many Landau levels are occupied for a given $B$. Thus, 
in terms of $\sigma_D$ and $g_B$, one can consider two distinct limits, i.e., weak field and the strong field. In the former case, $\mu_0 >>\frac{v_F}{l_B}$ which yields $\sigma_D \approx \frac{e^2\mu_0^2\tau}{12\pi^2 v_F}$, $g_B\approx \frac{\mu_0^2}{2\pi v_F^3}$ and  $\omega_p^2\approx \frac{e^2\mu_0^2}{3\pi v_F} + \frac{16\pi^2 v_F^3 B^2}{\mu_0^2}(\frac{e^2}{4\pi^2})^2\approx \frac{e^2\mu_0^2}{3\pi v_F}$. On the contrary, in the latter case, $\mu_0 <<\frac{v_F}{l_B}$ which implies $\sigma_D \rightarrow 0$, $g_B\approx \frac{1}{4\pi^2 v_F l_B^2}$ and $\omega_p^2\approx 8\pi v_F e B(\frac{e^2}{4\pi^2})$. Here, $l_B$ is the magnetic length defined as, $l_B=\frac{1}{\sqrt{eB}}$. 
 
It is well known that for $\omega >\omega_p$, the wave number is real and the wave propagates without any attenuation. On the otherhand, for $\omega <\omega_p$, $k$ becomes purely imaginary and the wave gets attenuated. In the present case, the expression for the plasma frequency in the weak magnetic field case has two terms, the first term corresponds to the 
plasma frequency of a conventional nodal metal whereas the second term which is due to CME, modifies it in a Weyl semimetal. It should be noted here that the plasma frequency $\omega_p$ is directly proportional to the chemical potential which is in accordance with the earlier work\cite{sarma09,lv13}. In case of strong magnetic field, the plasma frequency is directly proportional to the squareroot of the magnetic field only. {\it This result can be regarded as a smoking-gun experimental signature in WSM}. Further, considering the parameters namely, $\mu_0\sim $ few $[meV]$, $v_F\sim 10^7[cm s^{-1}$] and B$\sim$ few $[Gauss]$, $\omega_p$ in the weak field limit is estimated to be in $[meV]$ 
range which is much less as compared to the conventional metals(plasma frequency in the conventional metals are in the range of $[eV]$). The effect of chiral anomaly on the 
plasma mode in both the intrinsic and doped WSM has been recently investigated microscopically in random phase approximation(RPA)\cite{zhou16} and it has been shown 
that the chiral anomaly leads to unconventional plasmon mode giving rise to frequency in 
the $[meV]$ range. In the case of strong field, $\omega_p$ is estimated to be $\sim$ $100 [meV]$ for $B$ $\sim$ few $[Tesla]$. Thus, our result, eventhough is purely classical, agrees well with that 
of a microscopic theory if one neglects the logarithmic corrections. Due to the smallness of the plasma frequency up to a field of few Tesla in WSM, the dielectric 
function as well as the refractive index ($\epsilon=n^2={1-\frac{\omega_p^2}{\omega^2}}$) will be real and positive in the optical frequency range. {\it Since $\omega>\omega_p$, the material will look transparent similar to that of conventional alkali metals}. Such a novel property of these materials can invite potential applications.

So far, we have discussed the role of finite energy separation of nodes in a Weyl 
semimetal and its consequences in terms of macroscopic electrodynamics. This leads to CME which provides significant changes in the properties, namely, the frequency dependent 
skin depth, the capacitive transport, the plasma frequency as well as the refractive index respectively as compared to the conventional metals. The low frequency properties are controlled by a natural length scale in the system called the chiral magnetic length. Further, the plasma frequency in these materials is shown to be directly proportional to the squareroot of the magnetic field. Since the plasma frequency lies below the optical(visible) frequency, the WSM will look transparent. However, recently 
it has been shown in a two-band  lattice model that the Weyl nodes and chirality are not required to obtain CME while they remain crucial for the chiral anomaly\cite{chang16}. 
The CME, similar to anomalous Hall effect results directly from the Berry curvature of 
the energy bands even when there are no monopole source from the Weyl nodes. The 
phenomena of CME, thus can be observed in a wider class of materials.

\bigskip 
\noindent {\bf Acknowledgements:} \\ 
The author would like to thank Prof. T. V. Ramakrishnan and Prof. V. S. Subrahmanyam for stimulating discussions and critically reading the manuscript.


\begin{thebibliography}{00}
\bibitem{dsa1} Due to the axion action mentioned in the introduction, the Gauss's law and 
the Ampere's law can respectively be written as, $\vec{\nabla}.\vec{E}=4\pi
(\rho+\frac{\alpha}{2\pi^2}\vec{Q}.\vec{B}) $
and $\vec{\nabla}\times\vec{B}=\frac{4\pi}{c} [\vec{J} +\frac{\alpha}{2\pi^2}Q_0\vec{B}+\frac{\alpha}{2\pi^2} (\vec{Q}\times\vec{E})]+\frac{1}{c}\frac{\partial\vec{E}}{\partial t}$. Since we are 
interested here in CME, the terms containing $\vec{Q}$ have been neglected and the term 
related to $Q_0$ has been absorbed in $\sigma_{ch}$ as, $\frac{\alpha}{2\pi^2}Q_0=\sigma_{ch}$.
\bibitem{dsa2} The skin depth in a Weyl semimetal might develop an extra frequency 
dependence through the parameter $\sigma_{ch}$ in addition to the frequency dependence 
mentioned in the text which is beyond the scope of the present manuscript.
\end{thebibliography}
\end{document}